\documentclass[usenatbib]{mn2e}

\usepackage{amsmath}
\usepackage{graphicx}

\usepackage{savesym}
\usepackage{amsmath}
\savesymbol{iint}
\usepackage{txfonts}

\usepackage{subfigure}

\newcommand       \bea          {\begin{eqnarray}}
\newcommand       \eea          {\end{eqnarray}}
\newcommand       \apj          {ApJ}
\newcommand       \apjl         {ApJL}

\newcommand       \nat          {Nature}
\newcommand       \mnras        {MNRAS}

\newcommand{\be}{\begin{equation}}
\newcommand{\ee}{\end{equation}}

\def\simlt{\mathrel{\hbox{\rlap{\hbox{\lower4pt\hbox{$\sim$}}}\hbox{$<$}}}}
\def\gtrsim{\mathrel{\hbox{\rlap{\hbox{\lower4pt\hbox{$\sim$}}}\hbox{$>$}}}}
\def\lesssim{\mathrel{\hbox{\rlap{\hbox{\lower4pt\hbox{$\sim$}}}\hbox{$<$}}}}

\title[Diversity of Transients from Magnetar Birth]{The Diversity of Transients from Magnetar Birth}
\author[]{
Brian D.~Metzger$^{1}$,
Ben Margalit$^{1}$,
Daniel Kasen$^{2,3},$
Eliot Quataert$^{2}$\\
$^{1}$Columbia Astrophysics Laboratory, Columbia University, New York, NY, 10027, USA\\
$^{2}$Departments of Physics and Astronomy, Theoretical Astrophysics Center, University of California, Berkeley, CA, USA\\
$^{3}$Nuclear Science Division, Lawrence Berkeley National Laboratory, Berkeley, CA, USA\\
}

\begin{document}

\date{Received / Accepted}
\pagerange{\pageref{firstpage}--\pageref{lastpage}} \pubyear{2014}

\maketitle

\label{firstpage}

\begin{abstract}Strongly-magnetized, rapidly-rotating neutron stars are contenders for the central engines of both long-duration gamma-ray bursts (LGRBs) and hydrogen-poor super-luminous supernovae (SLSNe-I).  Models for typical ($\sim$ minute long) LGRBs invoke magnetars with high dipole magnetic fields ($B_{\rm d} \gtrsim 10^{15}$ G) and short spin-down times, while models for SLSNe-I invoke neutron stars with weaker fields and
longer spin-down times of weeks.  Here we identify a transition region in the space of $B_{\rm d}$ and birth period for which a magnetar can power both a long GRB and a luminous SN.  In particular, we show that a 2 ms period magnetar with a spin-down time of $\sim 10^{4}$ s can explain the observations of both the ultra-long GRB 111209 and its associated luminous SN2011kl.  For magnetars with longer spin down times, we predict even
longer duration ($\sim 10^{6}$ s) GRBs and brighter supernovae, a correlation that extends to Swift J2058+05 (commonly interpreted as a tidal disruption event).  We further show that previous estimates of the maximum rotational energy of a proto-magnetar were too conservative and energies up to $E_{\rm max} \sim 1-2\times 10^{53}$ ergs are possible.  The magnetar model can therefore comfortably accommodate the extreme energy requirements recently posed by the most luminous supernova ASASSN-15lh.  The high ionization flux from a pulsar wind nebula powering ASASSN-15lh may lead to an “ionization break-out” X-ray burst over the coming months, which would be accompanied by an abrupt change in the optical spectrum.  We conclude by briefly contrasting millisecond magnetar and black hole models for SLSNe and ultra-long GRBs.

 \end{abstract}

%\keywords{gamma-ray bursts: general}


\section{Introduction}
\label{sec:intro}
\vspace{-0.2cm}
In addition to canonical long-duration gamma-ray bursts (LGRBs) with characterstic durations of tens of seconds, a handful of bursts lasting longer than 1000 seconds have been discovered (\citealt{Levan+14}).   These ``ultra-long" GRBs (ULGRBs) share many properties with LGRBs of normal duration, including rapid variability, large isotropic radiated energies close to a solar rest mass, and an association with star-forming galaxies at cosmological distances (see \citealt{Levan15} for a review).  
%Members of this small but growing class include GRB 101225A (\citealt{Thone+11}; \citealt{Campana+11}), 111209A (\citealt{Levan+14}), 121027A (\citealt{Tanvir+12}) and 130925A (\citealt{Evans+14}; \citealt{Piro+14}).  
It remains debated whether ULGRBs are simply the longest lasting members of a single, continuous LGRB population (\citealt{Virgili+13}; \citealt{Zhang+14}) or whether they represent a distinct class with potentially different progenitors (\citealt{Boer+15}; \citealt{Levan+14}).  

Most LGRBs are accompanied by stripped envelope, hyper-energetic supernovae (SNe) (\citealt{Woosley&Bloom06}), making it natural to ask whether the same is true for ULGRBs.  The redshift of GRB 121027A was too high to search for a SN, while the afterglow of GRB 130925A was heavily extinguished by dust.  The afterglows of GRB 101225A and 111209A both showed blue to red color evolution several weeks after the burst, consistent with a SN brightening above the power-law afterglow (\citealt{Levan+14}; \citealt{Greiner+15}).  However, the spectrum of these thermal transients extended further into the ultraviolet than other GRB SNe.  The lack of clear spectral features was explained as Doppler line blending resulting from high ejecta velocities of $\gtrsim 20,000$ km s$^{-1}$ (\citealt{Greiner+15}).   

If ULGRBs result from the core collapse of massive stars then, as with other LGRBs, central engine models fall into two classes: a hyper-accreting stellar-mass black hole (BH) (\citealt{Woosley93}) or a neutron star (NS) with a millisecond rotation period and an ultra-strong magnetic field (a ``millisecond proto-magnetar"; \citealt{Usov92}).  In magnetar models the GRB jet is powered by the rotational energy extracted by electromagnetic torques, and hence the GRB is generally weaker and longer in duration for magnetars with weaker magnetic fields.  A BH engine, by contrast, is powered only so long as it continues to accrete stellar debris.  The durations of ULGRBs challenge BH models because the majority of the bound debris returns on the gravitational free-fall time of the stellar envelope (e.g., \citealt{Perna+14}).  This is usually only minutes for the compact Wolf-Rayet progenitors of normal LGRBs, but can be months or longer for blue super-giants (\citealt{Quataert&Kasen12}; \citealt{Woosley&Heger12}; \citealt{Gendre+13}).  However, the lack of hydrogen spectral features in the SNe which accompany ULGRBs may disfavor such extended stellar progenitors (\citealt{Greiner+15}).      

A long-lived central engine is also a leading explanation for the hydrogen-poor class of ``super-luminous supernovae" (SLSNe Type I, or SLSNe-I), stellar explosions with peak luminosities 10 to 100 times higher than normal core collapse SNe  (\citealt{Quimby+11}; \citealt{GalYam12}) and which cannot be powered by the radioactive decay of $^{56}$Ni. 
% Some luminous SNe are powered by shock interaction between the SN ejecta and a dense, hydrogen-rich circumstellar medium (e.g., \citealt{Smith+07}) and this represents a possible explanation for some SLSNe-I as well (\citealt{Moriya+10}; \citealt{Chevalier&Irwin11}).
  In engine-powered models for SLSNe-I, the central compact object is again either a millisecond proto-magnetar (\citealt{Kasen&Bildsten10}; \citealt{Woosley10}) or an accreting BH (\citealt{Dexter&Kasen13}).  The long-lived nature of the central engine, rather than just its total energy budget, is critical to enhancing the SN luminosity: energy released too early is degraded by adiabatic expansion of the ejecta before being radiated.  A connection between SLSNe-I and LGRBs is suggested by similarities in their host galaxy properties (\citealt{Lunnan+14};   \citealt{Leloudas+15}).

The discovery that SN 2011kl following GRB 111209A was significantly over-luminous as compared to normal GRB-SNe provides a potential link between ULGRBs and SLSNe-I, leading \citet{Greiner+15} to suggest a magnetar origin for this event.  This interpretation, however, also raises many questions:  If a magnetar (or an accreting BH) can separately produce both an LGRB and a SLSNe-I, can one produce both simultaneously?  Why are the SNe that accompany ULGRBs different than the optically-selected SLSNe-I?  Why are GRB-SNe usually not as luminous as SN 2011kl?

Adding fuel to the debate was the recent discovery of ASASSN-15lh, a SLSNe-I with the highest peak luminosity ($\gtrsim 2\times 10^{45}$ ergs s$^{-1}$) of any SN known to date.  Its total radiated energy of $\approx 10^{52}$ erg represents a significant fraction of the maximum rotational energy of a NS, posing a challenge to the magnetar model for SLSNe-I (and, indeed, to any model).  In a similar way, the gamma-ray and afterglow kinetic energies of the brightest LGRBs detected by {\it Fermi} are inferred to approach or exceed a few 10$^{52}$ ergs (\citealt{Cenko+11}).  This apparent fine tuning of the spin rate so close to its maximum has generally been used to argue against the magnetar model for LGRBs, although the clustering of GRB-SNe kinetic energies near the same scale has been used to argue {\it in favor} of a magnetar engine (\citealt{Mazzali+14}). 

In light of GRB 111209A/SN2011kl and ASASSN-15lh, as well as the rapidly growing sample of SLSNe-I (e.g., \citealt{Inserra+13}, \citealt{Nicholl+14}), we revisit the range of observational signatures of magnetar birth, placing GRBs and SLSNe-I within a common framework.  We show that ``transition" events like LGRB 111209A/SN2011kl are a natural consequence of the model, provided that a relativistic jet can escape the star.  Furthermore, although the energy of ASASSN-15lh was extreme, we show that previous estimates of the maximum energy of a magnetar have been underestimated.
%: SLSNe or LGRBs several times more energetic than ASASSN-15lh are in principle possible.

%This paper is organized as follows.  In $\S\ref{sec:magnetar}$ we briefly review the magnetar model for GRBs and SLSNe-I.  

\vspace{-0.5cm}

\section{Energetics of Proto-Magnetar Spindown}
\label{sec:magnetar}

The rotational energy of a NS of gravitational mass $M_{\rm ns}$ is given by
\be
E_{\rm rot} = I\Omega^{2}/2 \simeq 2.5\times 10^{52}(M_{\rm ns}/1.4M_{\odot})^{3/2}P_{\rm ms}^{-2}\,{\rm erg},
\label{eq:Erot}
\ee 
where $P = 2\pi/\Omega = P_{\rm ms}\,{\rm ms}$ is the rotational period and $I \simeq 1.3\times 10^{45}(M_{\rm ns}/1.4 M_{\odot})^{3/2}\,{\rm g\, cm^{2}}$ is the NS moment of inertia (\citealt{Lattimer&Schutz05}).  The maximum allowed rotational energy, $E_{\rm max}$, corresponds to the minimum spin period, $P_{\rm min}$, set by the mass-shedding limit.  In $\S\ref{sec:Emax}$ we show that $P_{\rm min}$ typically varies from $\approx 1$ ms for $M \approx 1.4M_{\odot}$ to 0.7 ms for $M \approx 2M_{\odot}$ (depending on the NS equation of state), such that $E_{\rm max}\gtrsim 10^{53}$ erg for massive NSs.

The pulsar loses rotational energy at a rate which, for an aligned force-free wind, is given by (\citealt{Contopoulos+99})
\begin{eqnarray}
L_{\rm sd} = \frac{\mu^{2}\Omega^{4}}{c^{3}} \simeq 1.7\times 10^{50}B_{15}^{2}P_{\rm ms}^{-4}\left(1 + \frac{t}{t_{\rm sd}}\right)^{-2}{\rm erg\,s^{-1}} 
\label{eq:Lsd}
\end{eqnarray}
where $\mu = B_{\rm d}R_{\rm NS}^{3}$ is the dipole moment, $B_{\rm d} = 10^{15}B_{15}\,{\rm G}$ is the surface equatorial dipole field\footnote{Note that our definition of $B_{\rm d}$, which agrees with that most commonly used in the pulsar community, is lower than that value adopted by \citet{Kasen&Bildsten10} by a factor of $\sqrt{12} = 3.46$.}, $R_{\rm NS} = 12\,{\rm km}$ is the NS radius, and
\be
t_{\rm sd} = \left.\frac{E_{\rm rot}}{L_{\rm sd}}\right|_{t = 0}\simeq 147\,{\rm s}\,(M/1.4M_{\odot})^{3/2}B_{15}^{-2}P_{\rm 0,ms}^{2}
\label{eq:tsd}
\ee
is the initial spin-down time.  The spin-down luminosity exceeds the value given by equation (\ref{eq:Lsd}) for several seconds after core bounce due to the neutrino-heated wind (\citealt{Thompson+04}; \citealt{Metzger+11}), but this correction is minor for ULGRBs.
%, who instead use the polar field strength and assume vacuum spin-down for a dipole inclination angle of 45 degrees.

\subsection{Powering Gamma-Ray Bursts}

The magnetar outflow, though approximately isotropic on small radial scales, is collimated into a narrow jet along the polar axis via the confinement provided by the surrounding stellar envelope (\citealt{Uzdensky&MacFadyen06}, \citealt{Bucciantini+07}).  This collimation may be facilitated by the anisotropic stress induced by the strong toroidal magnetic field in the nebula separating the magnetar wind from the surrounding star (\citealt{Bucciantini+09}).  It remains uncertain whether such a strong, ordered magnetic field is maintained in the face of non-axisymmetric instabilities (\citealt{Begelman98}; \citealt{Porth+13}, \citealt{Mosta+14}).  

The formation of a stable jet does not guarantee that the jet will escape the star over the timescale $\approx t_{\rm sd}$ of peak spin-down power.  A jet of luminosity $L_{\rm j}$ and half-opening angle $\theta_{\rm j}$ requires a time $t_{\rm esc} \approx M_{\rm ej}v_{\rm ej}c\theta_{\rm j}^{2}/L_{\rm j}$ to escape the stellar ejecta of mass $M_{\rm ej}$ and velocity $v_{\rm ej}$.  The condition that $t_{\rm esc} < t_{\rm sd}$ for $L_{\rm j} \approx L_{\rm sd}|_{t = 0} \approx E_{\rm rot}/t_{\rm sd}$ translates into a maximum opening angle for jet escape,
\be
\theta_{\rm j} \lesssim \theta_{\rm j,max} \equiv \left(\frac{E_{\rm rot}}{M_{\rm ej}v_{\rm ej}c}\right)^{1/2} \approx \left(\frac{v_{\rm ej}}{2c}\right)^{1/2},\,\,\,\text{ jet escape}
\label{eq:thetaj}
\ee
(\citealt{Quataert&Kasen12}) where the second equality assumes that $E_{\rm rot} \approx M_{\rm ej}v_{\rm ej}^{2}/2$, as applies if the kinetic energy of the explosion is dominated by the magnetar rotational energy.  Condition (\ref{eq:thetaj}) is not easily verified by current simulations, but it still provides a consistency check on the opening angle derived by modeling ($\S\ref{sec:Greiner}$).   

The power of a jet that cleanly escapes the star will approximately equal the peak spin-down power of the NS, i.e. $L_{\rm j} \approx L_{\rm sd}|_{t = 0}$ (eq.~[\ref{eq:Lsd}]) (\citealt{Bucciantini+07}).  The observed isotropic $\gamma-$ray luminosity is then $L_{\rm iso} \approx \epsilon_{\gamma}f_{\rm b}^{-1}L_{\rm sd}|_{t = 0}$, where $f_{\rm b} \approx \theta_{\rm j}^{2}/2$ is the beaming fraction and $\epsilon_{\gamma}$ is the radiative efficiency.  The LGRB duration is most naturally associated with the spin-down time over which $L_{\rm sd}$ is roughly constant, i.e. $T_{\gamma} \approx t_{\rm sd}$ (eq.~[\ref{eq:tsd}]), assuming the jet propagation time through the star is negligible in comparison.  

\vspace{-0.5cm}

\begin{figure*}
\subfigure{
\includegraphics[width=0.74\textwidth]{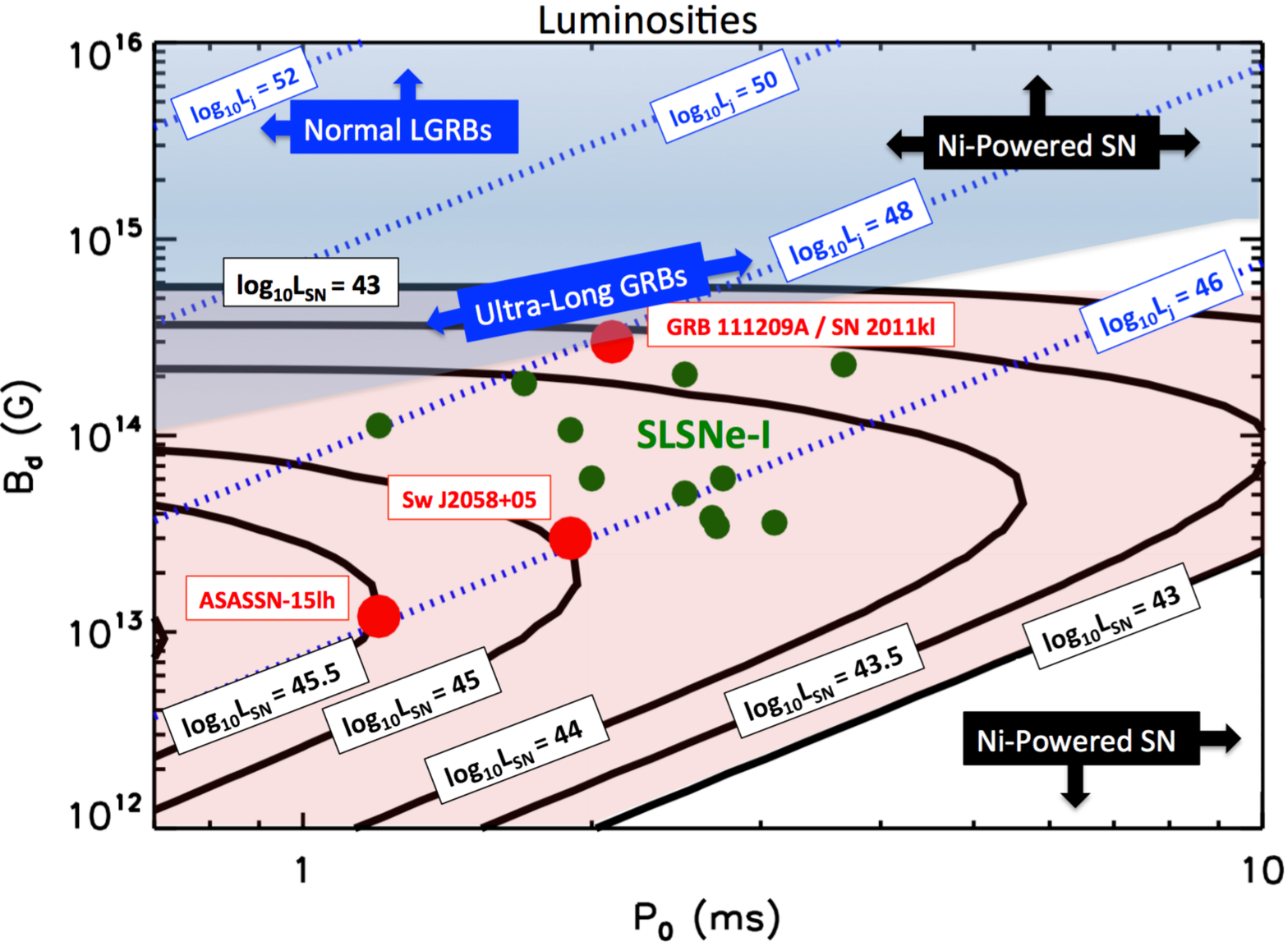}}
\subfigure{
\includegraphics[width=0.75\textwidth]{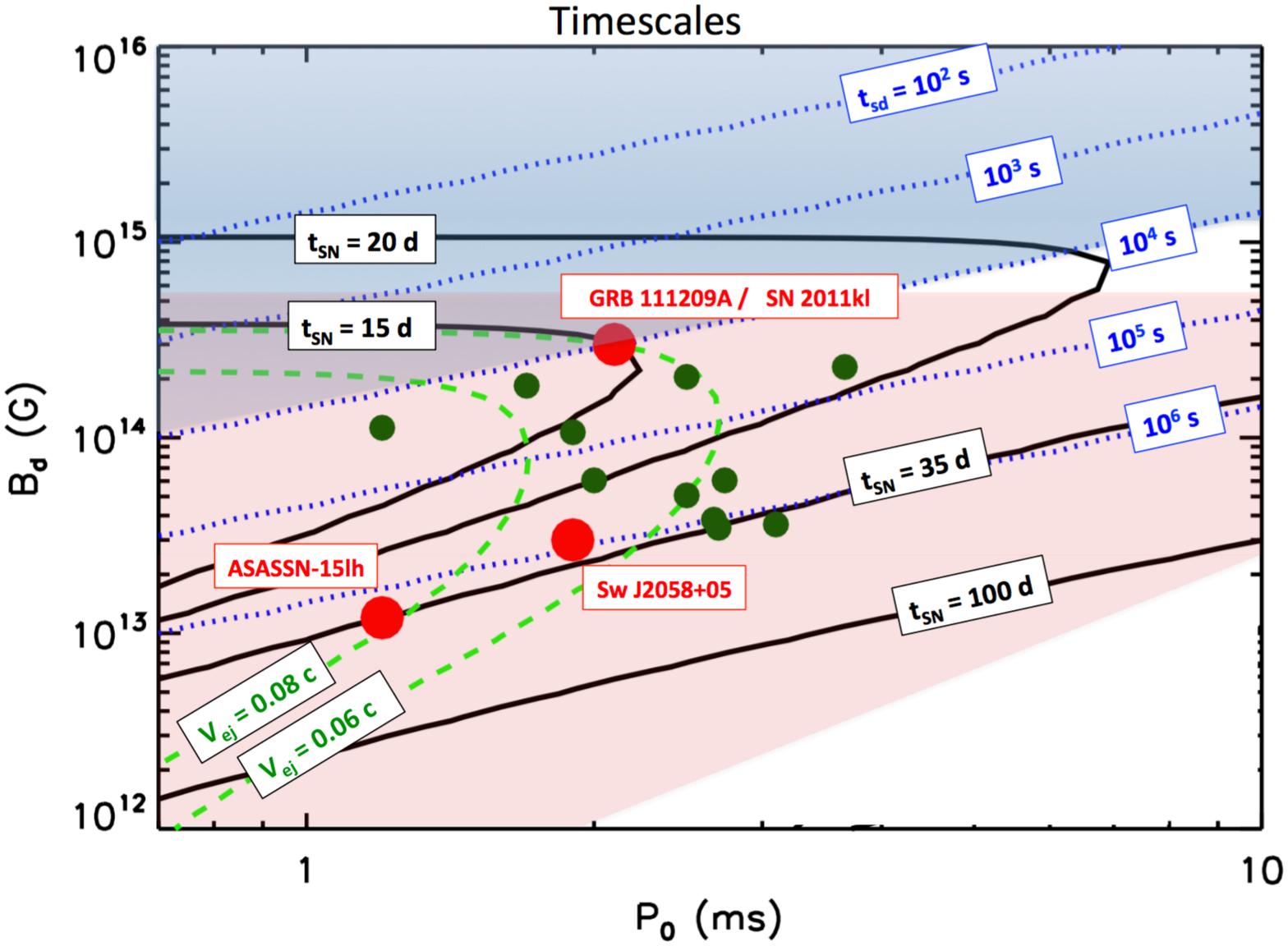}} 

\vspace{-0.4cm}
\caption{Parameter space of dipole magnetic field $B_{\rm d}$ and birth spin period $P_{0}$, calculated for $M_{\rm ej} = 3 M_{\odot}$, $M_{\rm Ni} = 0.2M_{\odot}$, $\kappa = 0.2$ cm$^{2}$ g$^{-1}$, $E_{\rm k,0} = 10^{51}$ erg and $M = 1.5M_{\odot}$.  Dashed blue contours show the peak spin-down luminosity (top panel) and the characteristic spin-down timescale (bottom panel), the later potentially associated with the LGRB duration.  Solid black contours show the peak SN luminosity $L_{\rm SN}$ (top panel) and peak timescale $t_{\rm SN}$ (bottom panel).  Superluminous (SLSNe), powered by NS rotational power, have $L_{\rm SN} \gtrsim 10^{43}$ erg s$^{-1}$.  Dark green circles show the NS parameters inferred by previous fits to other SLSNe-I by \citet{Nicholl+14} and \citet{Chatzopoulos+13} (we caution that the best fit ejecta mass was not usually equal to the 3$M_{\odot}$ assumed in calculating the contour lines).  Red circles show fits from this paper to GRB 111209A/SN2011kl (\citealt{Greiner+15}), ASASSN-15lh (\citealt{Dong+15}), and {\it Sw} J2058+05 (\citealt{Pasham+15}).  GRB111209A/SN2011kl is constrained by the requirements: SN peak luminosity $L_{\rm SN} = 3\times 10^{43}$ erg s$^{-1}$ (solid black line), SN peak time $t_{\rm SN} = 15$ days (dashed black line), ejecta velocity $v_{\rm ej} \approx 2\times 10^{4}$ km s$^{-1}$ (solid green line), and a GRB of duration $T_{\gamma} = t_{\rm sd} \approx 1.4\times 10^{4}$ s (dashed blue line).   ASASSN-15lh is constrained by: SN peak luminosity $L_{\rm SN} = 3\times 10^{45}$ erg s$^{-1}$ and SN peak time $t_{\rm SN} \approx 35$ days.  }
%A dashed brown line delineates the region above which the nebula inflated by the NS inside the exploding star can produce a shock break-out signal prior to the main SN peak (\citealt{Kasen+15}).  Below the orange line the electron scattering optical depth of the ejecta at the time of the SN peak is $\lesssim 30$ ($\S\ref{sec:discussion}$).}
\label{fig:fits}

\vspace{-0.3cm}
\end{figure*}

\subsection{Powering Supernova Light Curves}
\label{sec:SLSN}

The SN explosion ejects a mass of $M_{\rm ej}$ with an initial kinetic energy $E_{\rm 0,k} = 10^{51}$ erg.  As the ejecta expands, its thermal energy $E$ evolves as a function of time $t$:
\be \frac{dE}{dt} = -\frac{E}{R_{\rm ej}}v_{\rm ej}- L_{\rm SN} + L_{\rm Ni}+ L_{\rm sd}, 
\label{eq:dEdt}
\ee
where the first term accounts for PdV losses, $R_{\rm ej}$ and $v_{\rm ej} = dR_{\rm ej}/dt$ are the mean radius and velocity of the ejecta, respectively.  The kinetic energy $E_{\rm k} = M_{\rm ej}v_{\rm ej}^{2}/2$, increases as the result of PdV work, $dE_{\rm k}/dt = (E/R_{\rm ej})v_{\rm ej}$, as required by energy conservation.

The term $L_{\rm SN} = E/t_{\rm d}$ is the radiated luminosity, where $t_{\rm d} = (3\kappa M_{\rm ej})/(4\pi c v_{\rm ej}t)$ is the photon diffusion timescale, $\kappa$ is the optical opacity, and $
L_{\rm Ni} \approx 1.3\times 10^{43}\,{\rm erg\,s^{-1}}\,\left(M_{\rm Ni}/0.2M_{\odot}\right)\exp\left(-t/8.8{\rm d}\right)$ is the heating rate due to $^{56}$Ni decay, where $M_{\rm Ni}$ is the $^{56}$Ni mass.  The latter varies from $M_{\rm Ni} \approx 0.05-0.1M_{\odot}$ in normal Type Ib/c SNe to $M_{\rm Ni} \approx 0.3-0.5M_{\odot}$ in GRB SNe (\citealt{Nomoto+03}).  The opacity of $\kappa = 0.1-0.2$ cm$^{2}$ g$^{-1}$ is set by electron scattering and Doppler-broadened lines.  The last term in equation (\ref{eq:dEdt}) accounts for energy input from the NS (eq.~[\ref{eq:Lsd}]), which we assume thermalizes with unity efficiency (at late times $t \gg t_{\rm sd}$; see $\S\ref{sec:discussion}$).

%The ejecta radiates the thermal energy with high efficiency once $L_{\rm SN}$ exceeds the PdV loss term, i.e. at the characterstic time, 
%\be
%t_{\rm d,0} = \left(\frac{3\kappa M_{\rm ej}}{4\pi c v_{\rm ej}}\right)^{1/2} \approx 33\,{\rm d}\,\,\left(\frac{\kappa}{0.1\,{\rm cm^{2}\,g^{-1}}}\right)^{1/2} \left(\frac{M_{\rm ej}}{5M_{\odot}}\right)^{1/2}\left(\frac{v_{\rm ej}}{10^{4}\,{\rm km\,s^{-1}}}\right)^{-1/2},
%\label{eq:tpeak}
%\ee
%when the expansion time $t_{\rm exp} \approx r/v_{\rm ej}$ equals the diffusion time $t_{\rm d}$.  The value of $t_{\rm d,0}$ sets the characteristic rise time of the light curve.

\vspace{-0.5cm}

\section{The Diversity of Transients from Magnetar Birth}

We survey the landscape of GRB/SN transients produced by millisecond NS birth for a range of dipole fields $B_{\rm d} \sim 10^{12}-10^{16}$ G and initial spin periods $P_{0} \simeq 0.7-10$ ms.   All solutions are calculated for $M_{\rm ej} = 3 M_{\odot}$, $M_{\rm Ni} = 0.2M_{\odot}$, $\kappa = 0.2$ cm$^{2}$ g$^{-1}$, and $M = 1.5M_{\odot}$.  
% maximum rotational energy of $E_{\rm max} \approx 2.5\times 10^{52}$ ergs ($P_{\rm min} \approx 1$ ms; Fig.~\ref{fig:Erot}), but we nevertheless consider shorter rotational periods to account for for more massive NSs.

Figure \ref{fig:fits} overviews the $B_{\rm d}-P_{0}$ plane, with the top and bottom panels showing contours of GRB/SN luminosities and peak timescales, respectively.  Dashed blue lines show the maximum jet luminosity $L_{\rm j} = L_{\rm sd}|_{t = 0}$ (top panel; eq.~[\ref{eq:Lsd}]) and the initial spin-down time $t_{\rm sd}$ (bottom panel; eq.~[\ref{eq:tsd}]).  Magnetars in the far upper-left corner of the diagram produce high jet luminosities and short spin-down times $t_{\rm sd} \lesssim 100$ s, characteristic of the (beaming-corrected) luminosities and durations of normal LGRBs.  Moving to longer birth periods and lower $B_{\rm d}$, the spin-down time increases and the jet becomes weaker.  If the jet is stable and can escape the star for $t_{\rm sd} \gtrsim 10^{3}-10^{4}$ s, such events could produce ULGRBs.

Solid black lines show the peak SN luminosity, $L_{\rm SN}$ (top panel), while dashed lines show the time of the SN peak, $t_{\rm SN}$ (bottom panel).  For both the highest magnetic fields of $B_{\rm d} \gtrsim 10^{15}$ G, as well as very low $B_{\rm d}$/large $P_{0}$, which likely characterize the birth of ordinary pulsars, the SN light curve is predominantly powered by $^{56}$Ni with a characteristic duration of $t_{\rm SN} \approx 30$ days.  However, for $P \lesssim 10$ ms and intermediate values of $B_{\rm d} \sim 10^{12}-10^{15}$ G the SN luminosity is enhanced to a superluminous level by energy input from the pulsar (\citealt{Kasen&Bildsten10}).  The ``sweet spot" at $B_{\rm d} \approx 10^{13}$ G and $P_{0} \approx 1$ ms corresponds to a NS containing as much rotational energy as possible, with a spin-down time $t_{\rm sd}$ comparable to the ejecta diffusion time at peak light.  Green points show the best-fit magnetar parameters for other SLSNe-I (\citealt{Nicholl+14}; \citealt{Chatzopoulos+13}), while red points show the fits from this paper for GRB111209A/SN2011kl, ASASSN-15lh, and {\it Swift} J2058+05. 

Figure \ref{fig:fits} shows that the region of normal duration LGRBs overlaps with the Ni-powered SN regime, consistent with the standard interpretation that GRB-SNe are powered by radioactive Ni (as in other Type Ic SNe; e.g., \citealt{Nomoto+03}).  However, ULGRBs occupy a region of parameter space that overlaps that of SLSNe-I, especially for magnetars with short birth periods of $P_0 \lesssim 2$ ms.  Some of the longest GRBs could thus be accompanied by very luminous SNe \citep{Quataert&Kasen12}.

\vspace{-0.5cm}
\subsection{GRB 111209A/SN 2011kl}
\label{sec:Greiner}

%\begin{figure}
%\includegraphics[width=\textwidth] {ASASSN_LC.eps}
%\caption{Best-fit magnetar model ($B_{\rm d} = 1.2\times 10^{13}$ G; $P_{0} = 1.2$ ms) to the U-band light curve of ASASSN-14lh (blue circles; \citealt{Dong+15}).}
%\label{fig:ASASSN_LC}
%\end{figure}

\begin{figure}
\includegraphics[width=0.5\textwidth] {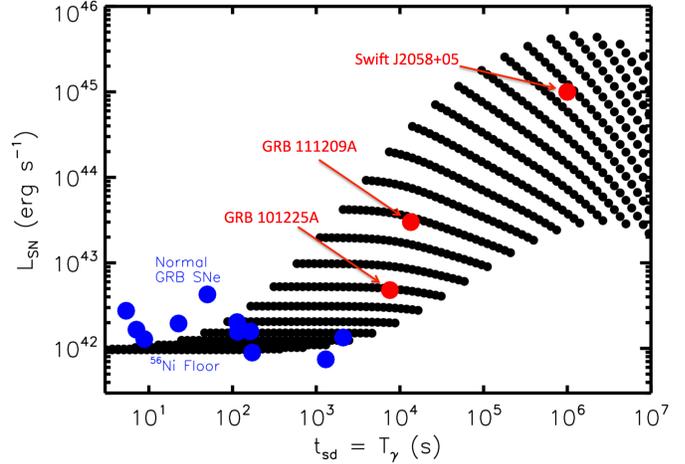}

\vspace{-0.5cm}
\caption{Peak SN luminosity, $L_{\rm SN}$, as a function of the magnetar spin-down time $t_{\rm sd}$ (eq.~[\ref{eq:tsd}]) for a range of models (black circles) with $B_{\rm d} \in [10^{12},10^{16}]$ G and $P_{\rm 0} \in [1,10]$ ms.  For short spin-down times, rotational energy is injected too early to significantly increase the SN luminosity above its minimum value set by $^{56}$Ni heating (which can vary moderately from the value of $M_{\rm Ni} = 0.2M_{\odot}$ assumed here).  However, for longer $t_{\rm sd}$ the maximum allowed value of $L_{\rm SN}$ increases.  Assuming that the prompt GRB duration approximately equals the spin-down timescale, i.e. $T_{\gamma} \approx t_{\rm sd}$, we show for comparison normal LGRB-SNe (blue circles; \citealt{Cano13}) and the ULGRBs (red circles) 111209A (\citealt{Greiner+15}), 101225A (\citealt{Levan+14}), and Swift J2058+05 (\citealt{Pasham+15}). }
\label{fig:LpkT90}

\vspace{-0.5cm}
\end{figure}

SN 2011kl peaked at a luminosity of $L_{\rm pk} \approx 3\times 10^{43}$ erg s$^{-1}$ on a timescale of 14 days (\citealt{Greiner+15}).  This fast rise, coupled with the high ejecta velocity ($\gtrsim 20,000$ km s$^{-1}$) required to explain the spectrum, requires a low ejecta mass of a few solar masses.  Figure \ref{fig:fits} shows that GRB 111209A and SN 2011kl are simultaneously explained by a magnetar with $B_{\rm dip} \approx 3\times 10^{14}$ G and $P \approx 2$ ms for $M_{\rm ej} \approx 3M_{\odot}$.  The preferred model has two free parameters (for fixed $M_{\rm ej}$) and lies at the intersection of four constraints: the peak duration and luminosity of the SN (dashed and solid black lines), the mean velocity of the ejecta (green line), and the GRB duration of $T_{\gamma} \approx t_{\rm sd} \approx 1.4\times 10^{4}$ s (dashed blue line).  

The model then {\it predicts} a beaming-corrected jet luminosity of $L_{\rm j} \approx 10^{48}$ erg s$^{-1}$ (solid blue line).  The observed average isotropic gamma-ray luminosity of $L_{\gamma} \approx 4\times 10^{50}$ erg s$^{-1}$ (\citealt{Greiner+15}) then implies a jet opening angle of $\theta_{\rm j} \approx (2L_{\rm j}\epsilon_{\gamma}/L_{\gamma})^{1/2} \approx 0.05$, assuming a typical value of $\epsilon_{\gamma} = 0.5$ (\citealt{Zhang+07}).  This is consistent with the maximum opening angle of $\theta_{\rm j,max} \approx 0.17$ for which the jet is capable of escaping the star (eq.~[\ref{eq:thetaj}], assuming $v_{\rm ej} \approx 0.06$ c).  If the ejecta mass had been closer to the average value of $\approx 10M_{\odot}$ inferred for SLSNe-I (\citealt{Nicholl+15b}), then the jet opening angle for GRB 111209A would be much closer to the critical value.  That a successful jet emerged from this explosion might therefore be the result of its lower than average ejecta mass.  
%The opening angle we infer is also broadly consistent with that inferred from the possible jet break observed by \citet{Greiner+15} at $t_{\rm b} = 9.12 \pm 0.48$ days, for which $\theta_{\rm j} = 0.18(t_{\rm b}/9{\rm d})^{3/8}(E_{\rm K,iso}/10^{54}\,{\rm erg})^{-1/8}(n/{\rm cm^{3}})^{1/8}$, where $E_{\rm K,iso}$ is the isotropic jet kinetic energy and $n$ is the external density.  

Figures \ref{fig:fits} shows that ULGRBs from NSs with weaker magnetic fields or longer spin periods than that responsible for GRB 111209A should also produce dimmer SNe.  As shown in Figure~\ref{fig:LpkT90}, the maximum peak SN luminosity should correlate with the burst duration for large values of the latter.  Indeed, the SN associated with the ULGRB 101225A of shorter duration ($T_{\gamma} \approx 7\times 10^{3}$ s) had a peak absolute magnitude $M_B \approx -19$ (\citealt{Levan+14}), closer to normal GRB SNe.  Also following this trend is the extremely long X-ray transient {\it Swift} J2058+05 ($T_{\gamma} \sim 10^{6}$ s; \citealt{Pasham+15}), which, although generally interpreted as a tidal disruption event, could also be a core collapse event (\citealt{Quataert&Kasen12}).  A weaker correlation is expected for low values of $T_{\gamma}$ due to variations in the baseline luminosity set by the amount of $^{56}$Ni produced in the explosion.  However, it cannot be discounted that some normal LGRB-SNe could be at least partially engine-powered, lowering the required $^{56}$Ni mass.  The engine's contribution to the SN would also be underestimated by our model if the spin-down power is higher at early times than predicted by eq.~(\ref{eq:Lsd}) due, for instance, to a time-dependent NS magnetic dipole (or, in the case of BH-powered events, a complex accretion history).

\vspace{-0.5cm}

\subsection{ASASSN-15lh}
\label{sec:ASASSN}

ASASSN-15lh had a peak luminosity of $L_{\rm SN} \gtrsim 2\times 10^{42}$ erg s$^{-1}$ and a total radiated energy of $\approx 10^{52}$ ergs \citep{Dong+15}.   Figure \ref{fig:LpkT90} shows our best-fit ``magnetar" model for ASASS-14lh of $B_{\rm d} = 1.2\times 10^{13}$ and $P_{0} = 1.2$ ms, calculated again assuming an ejecta mass of $M_{\rm ej} = 3 M_{\odot}$.  The solution is constrained by fitting the peak luminosity and timescale of the SN, resulting in an ejecta velocity of $\approx 2.4\times 10^{8}$ cm s$^{-1}$.  It is challenging to power such a luminous event without also placing a comparable amount of kinetic energy into the ejecta via PdV work.  A large kinetic energy for ASASSN-15lh is suggested by the presence of a broad feature in the optical spectrum at 4200 $\AA$ with a width of 250 $\AA$, corresponding to a velocity spread of $v_{\rm ej} \approx 2\times 10^{4}$ km s$^{-1}$.

\vspace{-0.5cm}

\section{Maximum Neutron Star Rotational Energy}
\label{sec:Emax}

 Once accounting for both radiation and kinetic energy imparted to the ejecta, ASASSN-15lh requires an energy of $\gtrsim 2\times 10^{52}$ erg, close to the maximum rotational energy of a 1.4 $M_{\odot}$ NS.  Such fine tuning appears to push the NS model to its limit by requiring ASASSN-15lh to occupy a `pin prick' in $B_{\rm d}-P_{0}$ parameter space.  ASASSN-15lh-like events are, however, very rare: \citet{Dong+15} estimates a rate 10-100 times lower than average SLSNe-I, which are already only $\sim 10^{-4}$ of all core collapse SNe (\citealt{Quimby+11}).  Furthermore, previous estimates of the maximum rotational energy are too conservative when more massive NSs are considered. 
  
\begin{figure}
\subfigure{
\includegraphics[width=0.5\textwidth]{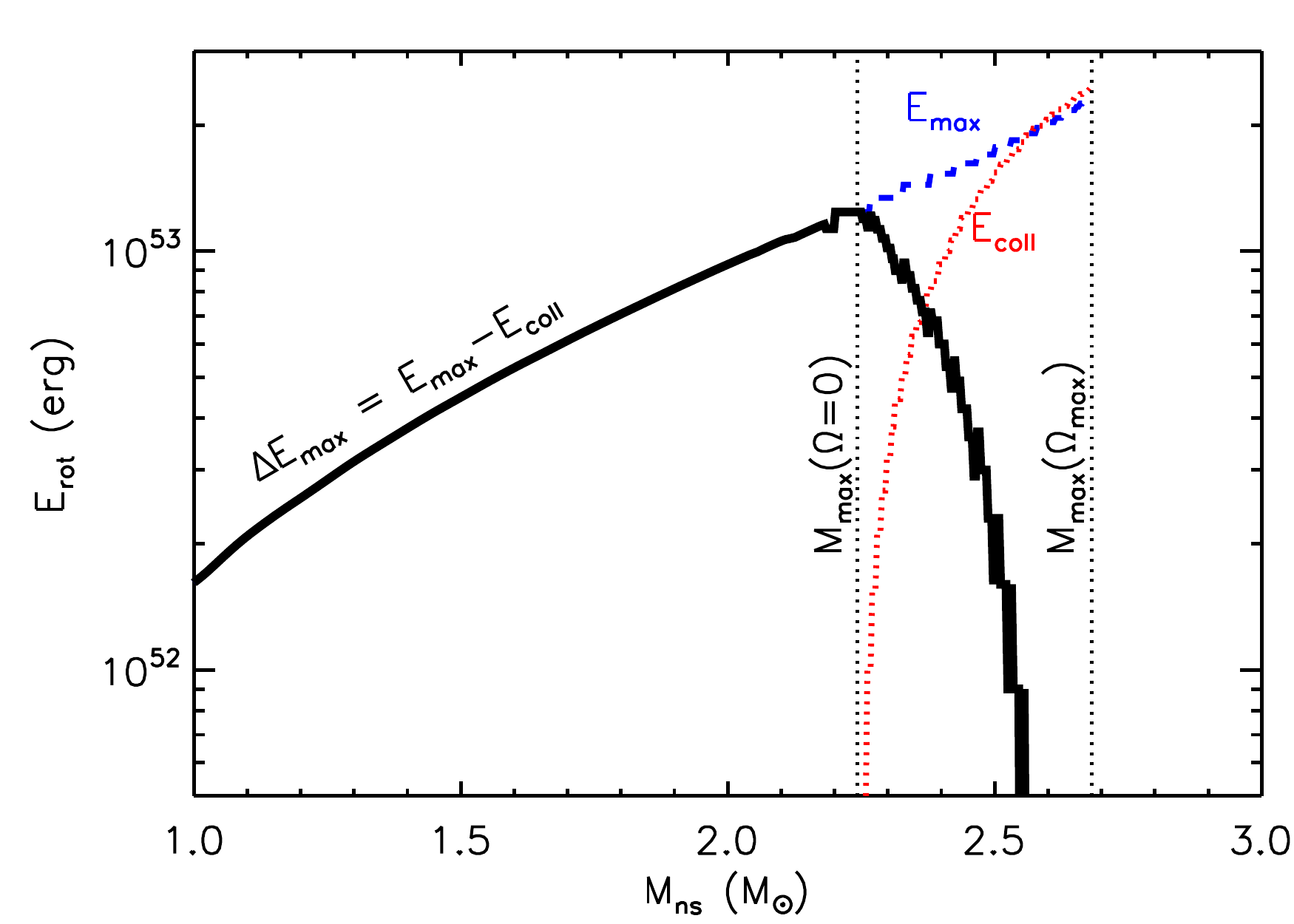}}

\vspace{-0.4cm}
\caption{Maximum extractable rotational energy from a NS, $\Delta E_{\rm max} \equiv E_{\rm max} - E_{\rm coll}$ (black solid line), as a function of the NS mass $M$, where $E_{\rm max}$ (blue dashed line) is the maximum rotational energy at the mass-shedding limit and $E_{\rm coll}$ (red dotted line) is the minimum rotational energy required for support of a supramassive NS against collapse to a BH.  The structure of the solid-body rotating NS is calculated using the {\tt rns} code assuming a parametrized piecewise polytropic EOS with an adiabatic index $\Gamma = 3$ above the break density of $\rho_{1} = 10^{14.7}$ g cm$^{-3}$ at a pressure of $P_{1} = 3.2\times 10^{34}$ dyn cm$^{-2}$.  We find that the maximum value of $\Delta E_{\rm max}$ lies within the relatively narrow range of $0.9-1.65 \times 10^{53}$ erg across a wide range of $\Gamma-P_{1}$ consistent with constraints on the maximum mass of a non-rotating NS.}
\label{fig:Erot}

\vspace{-0.3cm}
\end{figure}

Figure \ref{fig:Erot} shows the maximum allowed NS rotational energy $E_{\rm max}$ (black line) as a function of the gravitational mass, calculated  for a series of solid-body rotating NSs using the {\tt rns} code (\citealt{Stergioulas&Friedman95}).  We assume a parametrized piecewise polytropic EOS, as described in \citet{Margalit+15}, with an adiabatic index of $\Gamma = 3$ above the break density of $\rho_{1} = 10^{14.7}$ g cm$^{-3}$ and pressure $P_{1} = 3.2\times 10^{34}$ dyn cm$^{-2}$.  The particular EOS used in Fig.~\ref{fig:Erot} results in a radius of 10.6 km and gravitational mass of $M_{\rm max}(\Omega = 0) \approx 2.24M_{\odot}$.  Solid body rotation is established within minutes or less after NS formation due to the radial redistribution of angular momentum by magnetic stresses and gravitational waves. %(\citealt{Shapiro00}).

Figure \ref{fig:Erot} shows that a maximum rotational energy of $E_{\rm max} \approx 10^{53}$ erg is reached for a NS with $M \approx M_{\rm max}(\Omega = 0)$.  A solid-body rotating NS can possess a mass above this limit if it rotating sufficiently rapidly to be stabilized by centrifugal forces (a ``supramassive" NS).  Although the maximum rotational energy set by the mass-shedding limit continues to increase with $M > M_{\rm max}(\Omega = 0)$ along the supramassive branch, the {\it minimum} rotational energy needed to stabilize the NS from collapse, $E_{\rm coll}$ (red line), increases even faster with $M$.  Hence, the maximum available rotational energy, $\Delta E_{\rm max} = E_{\rm max}-E_{\rm coll}$ to be extracted decreases for $M > M_{\rm max}(\Omega = 0)$.  The peak of $\Delta E_{\rm max}$ thus occurs at $M \approx M_{\rm max}(\Omega = 0)$.  Although Figure \ref{fig:Erot} highlights our result for a particular EOS, we find that this maximum energy lies within the relatively narrow range of $0.9-1.65 \times 10^{53}$ erg across a wide range of EOS consistent with the lower limit set on the maximum mass of a non-rotating NS (\citealt{Antoniadis+13}).  This maximum rotational energy could in principle be limited by gravitational wave emission for the highest rotation rates.  However, we find that the ratio of kinetic energy to gravitational binding energy, T/W, of the mass-shedding sequence generally remains below the critical value of $(T/W)_{\rm max} \approx 0.13$ required for the growth of secular instabilities, driven by gravitational radiation reaction or viscosity (\citealt{Lai&Shapiro95}).  Thus, $E_{\rm max} \approx 10^{53}$ erg represents a new, relatively conservative upper limit on the energy permitted by the magnetar model, alleviating the tension produced by ASASSN-15lh and the most energetic LGRBs.
 
\vspace{-0.6cm}

\section{Discussion and Conclusions}

\label{sec:discussion}

Although ASASSN-15lh and GRB111209A/SN2011kl can be explained by the magnetar model, we have made several implicit assumptions.  A critical one is that the jet cleanly escapes the star at early times $\lesssim t_{\rm sd}$, placing a large fraction of its energy into the GRB jet, only to become trapped at later times $ \gg t_{\rm sd}$, such that the remaining rotational power thermalizes behind the ejecta with high enough efficiency to power the SN.  Such a transition is possible if the jet becomes less stable, or less effective at maintaining an open cavity through the expanding ejecta, as the spin-down power decreases from its peak initial value.  

It is useful to consider what conclusions can be drawn from the SLSNe-I population as a whole.  Although many of the current sample of SLSNe-I lie far from the ULGRB region (Fig.~\ref{fig:fits}), a few are sufficiently close to produce an ULGRB of duration $T_{\gamma} \gtrsim 10^{4}$ s should the jet escape the star.  Such high energy transients could easily have been missed due to relativistic beaming away from our line of site, especially given the beaming fraction inferred for GRB111209A of $f_{b}^{-1} \sim 400$ ($\S\ref{sec:Greiner}$).  Are the rates consistent? SLSNe-I occur at rates comparable to those of normal LGRBs (\citealt{Quimby+11}).  However, marginally over-luminous events like SN2011kl are more common than more luminous SLSNe-I, while ULGRBs may be intrinsically rare compared to normal GRBs, even after accounting for the bias against detecting lower luminosity events.  The SLSNe-I SCP06F6 did show an X-ray outburst of luminosity $L_{\rm X} \sim 10^{45}$ erg s$^{-1}$ several months after the explosion (\citealt{Levan+13}), but this may have signaled ionizing radiation escaping through the ejecta instead of a true jet (see below).  Orphan afterglow emission from an off-axis jet would peak at radio wavelengths on a timescale of months after the explosion, motivating a systematic radio search following nearby SLSNe.   

Figure \ref{fig:fits} also shows that the current SLSNe-I sample does not uniformly sample the NS parameter space capable of producing SLSNe, with a clear bias present for events with higher values of $B_{\rm d}$.  In principle, millisecond pulsars with fields as weak as $B_{\rm d} \approx 10^{12}$ G should produce SLSNe-I, albeit ones with slowly evolving light curves.  This dearth of low-$B_{\rm d}$ events may be intrinsic if NSs acquire strong magnetic fields as the result of a dynamo in the proto-NS phase caused by rapid rotation (\citealt{Thompson&Duncan93}).  It is also possible that assumptions of the SN model break down at low-$B_{\rm d}$, in particular regarding the efficiency with which the pulsar wind thermalizes its energy inside the SN ejecta.  As the ejecta expands and its density decreases, non-thermal radiation from the pulsar wind nebula can ionize its way through the ejecta (\citealt{Metzger+14}), allowing pulsar power to escape directly before thermalizing into optical radiation.  High energy gamma-rays also more readily escape at late times due to the decreasing $\gamma-\gamma$ and $\gamma-e$ cross sections at photon energies $\gg m_e c^{2}$.  If the thermalization efficiency is reduced for the low-$B_{\rm d}$ pulsars, which produce SLSNe-I that nominally peak at low Thomson optical depths $\tau_{\rm es} \lesssim 30$, this could help explain the apparent dearth of low-$B_{\rm d}$ NSs in the SLSNe-I population.

%We assume that the wind thermalizes with 100 percent efficiency at all times.  This may be a reasonable approximation near the time of the peak SN emission (\citealt{Metzger+14}), but becomes questionable at later times, for instance if high energy radiation escapes the ejecta without being absorbed and reradiated or downscattered to optical frequencies.  Thermalization may also be suppressed prior to the SN peak due to the high optical depth non-thermal radiation experience due to the presence of copious $e^{\pm}$ pairs (\citealt{Metzger&Piro14}; \citealt{Kasen+15}) or if energy leaks out of the nebula through the cavity carved by an escaping jet (i.e. that producing an earlier GRB).  We describe possible evidence for inefficient late-time thermalization based on the population of SLSNe-I in $\S\ref{sec:discussion}$.  

ASASSN-15lh is the only SLSNe-I that we find lies below $\tau_{\rm es}(t_{\rm SN}) < 30$.  Such luminous events can produce an ``ionization break-out" signal as soon as a few months after the SN (\citealt{Metzger+14}), characterized by an abrupt onset of soft X-ray emission of luminosity $L_{\rm X} \sim 10^{42}-10^{44}$ erg s$^{-1}$ and a drastic change in the optical spectrum to one characterized by nebular emission lines.  Such a break-out event is not guaranteed to occur (especially if the ejecta mass is high), but we strongly encourage high-cadence X-ray and optical spectroscopic coverage of ASASSN-15lh over the coming months, as its discovery would be a smoking gun for engine-powered models of SLSNe.  Other than their unusually long durations, ULGRBs resemble the bulk of the LGRB population in many of their other properties.  Therefore, if a single ULGRB can be unambiguously associated with magnetar birth, this strengthens the hypothesis that many (or all) LGRBs are magnetar-powered (\citealt{Thompson&Duncan93}; \citealt{Thompson+04}; \citealt{Metzger+11}; \citealt{Mazzali+14}).

%The magnetar model has recently been challenged by the discovery that some SLSNe-I are accompanied by early peaks in their light curves on timescales of days after the explosion, distinct from the main supernova peak occuring several weeks later (\citealt{Nicholl+15a}).  \citet{Kasen+15} proposed that a shock driven through the expanding SN ejecta by the high pressure of the nebula inflated by the magnetar could break-out from the stellar surface within days of the explosion, producing a luminous flare.  A minimum condition for a detectable break-out to occur is that the magnetar spin-down time remain active long enough for the break-out shock to become radiative.  This ``shock break-out" condition is shown as a brown line in Figure \ref{fig:main}, from which it is seen that this condition is satisfied by all of the SLSNe-I detected to date.  Break-out signals we not detected from most of these events, but could have been missed, either due to poorly-sampled time coverage of the early light curves (Smartt, private communication), or because the shock break-out peak was overwhelmed by the main SN emission.

Finally, much of the physics driving high energy transients and SLSNe from magnetar birth applies equally well if the engine is instead an accreting BH (\citealt{Dexter&Kasen13}), provided that the magnetar spin-down time is replaced by the period of peak accretion and the $t^{-2}$ decay is replaced by the commonly expected $t^{-5/3}$ decay in the mass fall-back rate from the stellar envelope.  Correlations between direct engine activity and SLSN luminosity such as those shown in Figure \ref{fig:LpkT90} should hold more generally.  Perhaps the best discriminant between NS and BH models for ULGRBs is the prediction of hydrogen in the SN ejecta in the BH case due to the requirement of an extended progenitor star.  Although hydrogen is not detected in the current sample of SNe accompanying ULGRBs, Doppler broadening of the lines could reduce their strength.  Future observations and more detailed modeling are thus required to place more meaningful constraints.  

If the jet power tracks the magnetic flux accumulated by the BH instead of just its accretion rate (\citealt{Tchekhovskoy&Giannios15}), a large progenitor star with a long free-fall time might not necessarily be required to explain ULGRBs.  In this case, LGRBs of shorter(longer) duration could result from stars with larger(smaller) magnetic flux.  However, in this case a seemingly fine-tuned correlation between the magnetic flux and the normalization of the fall back rate would be required to explain the similar total jet energies in normal and ULGRBs.

%An alternative models for ultra-long GRBs is the tidal disruption of a star by a massive black hole.  Indeed, a separate class of extremely long hard X-ray transients lasting $\sim 10^{6}$ s were interpreted to result from the tidal disruption of a star by a supermassive black hole (\citealt{Bloom+11}; \citealt{Levan+11}; \citealt{Burrows+11}; \citealt{Zauderer+11}).  Producing an event as luminous and as short as the $\sim 10^{3}-10^{4}$ s  duration of ultra-long GRBs is challenging from the disruption of a main sequence star because the fall-back timescale for the mostly tightly bound debris after a TDE is only $\lesssim 10^{4}$ seconds for a BH of mass $\lesssim 10^{3}$ s.  A faster timscale is possible following the tidal disruption of a white dwarf by an intermediate mass BH (\citealt{Krolik&Piran11}, \citealt{MacLeod+14}).  However, such a model would be challenged to explain the week-long thermal transient observed after the ultra-LGRBs (\citealt{Greiner+15}), since in most models for TDE thermal emission the thermal emission also peaks on the fall-back time.  

\vspace{-0.5cm}

\section*{ACKNOWLEDGMENTS}
We thank Jonathan Granot, Andrew Levan, Todd Thompson for helpful conversations, and Subo Dong for providing light curve data on ASASSN-15lh.  BDM and BM acknowledge support from the NSF grant AST-1410950 and the Alfred P. Sloan Foundation.  EQ was supported in part by NSF grant AST-1205732, a Simons Investigator award from the Simons Foundation and the David and Lucile Packard Foundation.  DK is supported in part by a Department of Energy Office of Nuclear Physics Early Career Award, and by the Director, Office of Energy Research, Office of High Energy and Nuclear Physics, Divisions of Nu- clear Physics, of the U.S. Department of Energy under Contract No. DE-AC02-05CH11231.

%\bibliographystyle{yahapj}
%\bibliography{ms}

\end{document}